\documentstyle[preprint,aps]{revtex}
\tightenlines
\begin{document}
\draft
\title{Second Harmonic Generation for a Dilute Suspension of Coated Particles} 
\author{P.M. Hui$^{1}$, C. Xu$^{2,1}$, and D. Stroud$^{3}$}
\address{$^{1}$ Department of Physics, The Chinese University of Hong Kong, \\
Shatin, New Territories, Hong Kong \\
$^{2}$ Department of Physics, Suzhou University, Suzhou 215006, People's Republic of China\\
$^{3}$ Department of Physics, The Ohio State University, Columbus, 
Ohio 43210-1106}
\maketitle

\begin{abstract}
We derive an expression for the effective 
second-harmonic coefficient of a dilute suspension of coated spherical
particles.  
It is assumed that the coating material, but not the core or the host,
has a nonlinear susceptibility for second-harmonic generation (SHG).
The resulting compact expression shows the various factors 
affecting the effective SHG coefficient. 
The effective SHG per unit volume of nonlinear coating
material is found to be greatly enhanced at certain frequencies,
corresponding to the surface plasmon resonance of the coated particles.
Similar expression is also derived for a dilute suspension of
coated discs.  For coating materials with third-harmonic (THG) 
coefficient, results for the effective THG coefficients are given 
for the cases of coated particles and coated discs. 

\end{abstract}

\bigskip
\newpage
\section{Introduction} \label{sec:intro} 

With the advancements in nanotechnology, it has
become possible to fabricate nanoparticles of various kinds with
specific geometries.  
For years, random or ordered composite materials consisting of
two or more materials with different physical properties have been
studied intensively, with the aim of
tuning the effective physical properties of the composite 
by properly choosing the constituent materials, the structure
of the composite, and/or the volume fraction of each of the constituents.
Ordered composites, for example, have led to the development
of semiconductor
heterostructures and devices, and photonic band-gap materials.
Composites consisting of small particles of one material
randomly embedded in a host medium also show interesting
behavior\cite{bergman1}.
The percolation effect that occurs as the volume fraction of one
component increases leads to a qualitative change in the physical response
of a random composite.  In the dilute concentration limit,
various physical responses are affected by local field effects due
to the inhomogeneous nature of the system. 

Of particular interest is the optical response of
nanoparticles and nanoparticles with specially designed
geometry\cite{shalaev1,flytzanis,liao,klar}.
For example, using a nanoshell geometry on particles, it has been found
that the effective Raman scattering can be enhanced by as much as
$10^{6}$ \cite{jackson}.  The surface plasmon resonance is found to be
shifted in nanoparticles coated with gold \cite{averitt}.
The optical properties of nanoparticles may also be affected by
quantum size effects\cite{zhou}.  
In general, it is expected that the local field effects play a
significant role in the nonlinear response in composites consisting of
small particles as the response depends on high powers of the
field\cite{shalaev2}.
Greatly enhanced nonlinear response in composite systems, when coupled
with a fast response time,  could lead to applications 
in the design of switching devices and
optical communcation systems. An excellent review on the nonlinear
optical properties of random media has been given by
Shalaev\cite{shalaev3}.
An up-to-date account on problems related to computational electromagnetics
in dielectric heterostructures has recently been given by
Brosseau and Beroual\cite{brosseau}. 
The current status on research on the properties
of nanostructured
random media is best reflected in the recent compilation by
Shalaev\cite{shalaev4}.

A number of theories has been developed for 
weakly nonlinear
composites\cite{shalaev3,stroud1,stroud2,neeves,zeng,haus,bergman2,sipe},
with the focus mainly on third-order 
nonlinear susceptibility such as the Kerr effect.
In previous works\cite{hui1,hui2}, 
we derived general expressions for nonlinear susceptibility
for the second harmonic generation (SHG) and third harmonic generation (THG) 
in random composites.  The authors also developed simple approximation
for the dilute concentration limit and effective-medium type approximation
for arbitrary concentrations.  With recent advancements in the fabrication
of coated particles and with possible enhanced local field effect
in mind, we study SHG in a dilute suspension of coated spheres.
Specifically, we assume that only the coating material has a
non-vanishing SHG coefficient.
The core and the host medium are linear in nature.
We derive an expression for the effective SHG coefficient per
unit volume of nonlinear material in the system.  It is found that
with modest choice of material parameters, the SHG coefficient
can be greatly enhanced at certain frequencies.  We also study the
two-dimensional (2D) version of the problem, i.e., the SHG coefficient
in a 2D dilute suspension of coated discs in a linear host, and similar
enhancement is found.  When the coated material has a non-vanishing
THG coefficient, we give the resulting expression for the effective
THG coefficient per unit volume of nonlinear material in both 3D and
2D random composites.

The plan of the paper is as follows.  In Sec.II, we derive the
effective SHG coefficient in a dilute suspension of coated spheres
in a linear host.  We illustrate the possible enhancement effect
by considering a model system of particles with
a metallic core and coated by a nonlinear material.  In Sec.III,
we give the corresponding results for a 2D dilute suspension of
coated discs.  Results for the effective THG coefficient in both 3D and 2D
cases are given in Sec.IV, together with a summary.

\section{Formalism} \label{sec:formal}

We consider the effective susceptibility for second-harmonic generation (SHG)
of a dilute suspension of {\em coated} particles embedded in a linear host
in both three dimension (3D) and two dimension (2D). 
In 3D, the system consists of coated spheres; while in 2D, the system  
consists of coated cylinders with aligned axes (or coated discs). 
The spheres (cylinders) are supposed 
to have inner radius $r_{1}$ and outer radius $r_{2}$.  
The core medium is supposed to 
have (possibly complex) dielectric constant $\epsilon_1$, 
the coating has dielectric constant $\epsilon_2$, 
and the host has dielectric constant $\epsilon_3$.
In addition, the coating, but neither the host nor the core, has a 
nonlinear susceptibility for second-harmonic generation.    
This susceptibility is, in general, a tensor of rank
three\cite{butcher} and is denoted
by $d_{ijk}(-2\omega;\omega,\omega)$ or $d_{ijk}(\omega,\omega)$, 
with each of the subscripts 
running over the three Cartesian indices.  
Note that the elements of $d$ may be complex.  
The total volume fraction of coating material in the 
composite is $f$, and $f << 1$ is assumed. 

Hui and Stroud\cite{hui1} have derived a general expression 
for the {\em effective} nonlinear SHG 
susceptibility $d_{ijk}^e(-2\omega;\omega,\omega)$
of a composite,
under the assumption that the nonlinearity is weak.
When applied to the system considered here, the result may be written in 
the form
\begin{equation}
d_{ijk}^e=fd_{\ell mn}
\langle K_{i\ell}(2\omega)K_{mj}(\omega)K_{nk}(\omega)\rangle.
\end{equation}
Here 
\begin{equation}
K_{nk}({\bf x},\omega) = \frac{E_n({\bf x},\omega)}{E_{0,k}(\omega)}
\end{equation}
is the possible local field enhancement factor giving 
the $n$-th component of the local electric field at position  
${\bf x}$ and frequency $\omega$ when the applied 
field $E_0$ is in the $k$-th direction at frequency $\omega$.  
$\langle\cdots\rangle$ denotes an average
over the volume of the {\em nonlinear shell} and the average
is to be calculated in the {\em linear} limit.  We use the
convention that repeated indices are summed over.

Equations (1) and (2), as well as all the results below, are also obtained
in the quasistatic approximation, according to which
the local electric field may be written as the negative gradient of a
scalar potential.  This approximation is most accurate when the particle
radii are small compared to both the wavelength of light in the medium and
the skin depth of metallic particles.  The quasistatic approximation 
neglects such effects as radiative losses, which may become important at
shorter wavelengths; it also omits diffuse scattering, as discussed
further in the last section of the paper.

We consider the 3D case of coated spheres in a linear host. 
In the dilute concentration limit where the interaction between 
particles can be neglected, 
the electric field in the coated shell can be calculated 
using standard electrostatics\cite{em}.  The electric field 
in the spherical shell ${\bf E}_{s}$, where the subscribe denotes the 
spherical shell with the coating medium, in the presence of
an external field $E_{0}\hat{z}$ is related
to the electric potential via 
\begin{equation}
{\bf E}_{s} = - \nabla \Phi_{s} 
\end{equation}
with 
\begin{equation}
\Phi_{s}(r, \theta) = -{\cal E}_{s} 
r\cos \theta + \frac{p_{s} \cos\theta}{r^2}, 
\end{equation}
where 
\begin{eqnarray}
{\cal E}_{s} & = & \Gamma_{s} E_0 \\
p_{s} & = & \lambda_{s} r_{1}^3 \Gamma_{s} E_0. 
\end{eqnarray}
Here $\Gamma_{s}$ and $\lambda_{s}$ are given by 
\begin{eqnarray}
\Gamma_{s} & = & \frac{3\epsilon_3}
{\epsilon_2+2\epsilon_3+2(\epsilon_2-\epsilon_3)\lambda_{s}/\mu_{s}} \\
\lambda_{s} & =  & \frac{\epsilon_1-\epsilon_2}{\epsilon_1+2\epsilon_2},
\end{eqnarray}
where $\mu_{s}$ is related to the ratio of the outer to inner radii
\begin{equation}
\mu_{s} = (r_{2}/r_{1})^3.
\end{equation}
The electric field in the coating medium 
can then be explicitly evaluated with the result
\begin{equation}
{\bf E}_{s} = {\cal E}_{s} \hat{z} + 
\frac{3(xz\hat{x}+yz\hat{y}+z^2\hat{z})-r^2\hat{z}}{r^5}p_{s},
\end{equation}
where $r$ is the distance from the center of the sphere to the 
point under consideration inside the spherical shell. 
Note that $p_{s}$ has dimensions of a dipole moment, and  
it is the dipole moment of the coated particle in the linear limit.
Comparing with results of a single dielectric sphere in a host, the 
coated sphere behaves as a sphere with an effective 
dielectric constant $\tilde{\epsilon}_{s}$ of the form 
\begin{equation}
\tilde{\epsilon}_{s}=\frac{1+2\lambda_{s}/\mu_{s}}
{1-\lambda_s/\mu_s}\epsilon_2 ,
\end{equation}
with $\lambda_{s}$ and $\mu_{s}$ given by Eqs.(8) and (9). 
It can be immediately seen from Eq.(11) that the surface plasmon
resonance of a coated particle can be tuned\cite{klar,averitt}
by the material parameters
$\lambda_{s}$ and the geometrical parameter $\mu_{s}$ of the particle.

To proceed and for the purpose of illustrating the possible 
enhancement in SHG, 
we will make the assumption that of the
possible components of $d$ in the shell, only $d_{iii}$ are nonzero and 
that these are all equal, i.e., $d_{iii} = d(-2\omega;\omega,\omega) 
\equiv d_{\omega,\omega}$ 
($i$ = $x$, $y$, or $z$). 
Depending on the symmetry of the nonlinear materials concerned, 
$d_{ijk}$ would have vanishing components for certain combinations 
of the indices. For example, one may also have $d_{ijk} \neq 0$ only
for $i \neq j \neq k$.  Our calculations can also be carried out
for the latter case. 
Using Eq.(1) in the dilute concentration limit gives
\begin{equation}
d^{e}_{\omega,\omega} = fd_{\omega,\omega}\sum_{i=x,y,z}
\langle K_{zi}(2\omega)K_{iz}^2(\omega)\rangle_{shell},
\end{equation}
where we have denoted the surviving SHG coefficient $d^{e}_{zzz}$
by $d^{e}_{\omega,\omega}$, for simplicity. 
The brackets $\langle\cdots\rangle_{shell}$ denote the average of the
enclosed quantity over the volume of the spherical shell, 
and $f$ is the volume fraction of coating material in the system. 

It is easily shown that only the $i = z$ term in the above sum contributes,
when a suitable angular average is carried out over the volume of the shell.
In order to evaluate that remaining term in Eq.(12), 
we need to calculate the average over the shell volume of the quantity
\begin{equation}
\left[{\cal E}_s(2\omega)+\frac{3z^2-r^2}{r^5}p_{s}(2\omega)\right]
\left[{\cal E}_s(\omega)+\frac{3z^2-r^2}{r^5}p_{s}(\omega)\right]^2,
\end{equation} 
where we have indicated the frequency at which the field and the
dipole moment are to be evaluated explicitly. 
Defining $B_{s} = (3z^{2}-r^{2})/r^{5}$,
the average is easily evaluated by noting that 
\begin{eqnarray}
\left\langle B_{s} \right\rangle_{shell} & = & 0 \nonumber \\
\left\langle B_{s}^{2} \right\rangle_{shell} & \equiv S_2 & =   
\frac{1}{v_{shell}}\left(\frac{1}{r_{1}^3}-\frac{1}{r_{2}^3}\right)
\frac{16\pi}{15} \nonumber \\
\left\langle B_{s}^{3} \right\rangle_{shell} & \equiv S_3 & = 
\frac{1}{v_{shell}}\left(\frac{1}{r_{1}^6}-\frac{1}{r_{2}^6}\right)
\frac{32\pi}{105}.
\end{eqnarray}
Here $v_{shell}=(4\pi/3)(b^3-a^3)$ is the volume of the spherical shell. 
Carrying out the average of expression (13), 
the effective SHG susceptibility is then given in terms of 
these quantities by 
\begin{eqnarray}
\frac{d^{e}_{\omega,\omega}}{f d_{\omega,\omega}} 
& = & \frac{1}
{E_0(2\omega)E_0^2(\omega)} 
\left( {\cal E}_s(2\omega){\cal E}_s^{2}(\omega)  \right. \nonumber \\
& & \left. + S_{2}\left[{\cal E}_s(2\omega)p_{s}^2(\omega) 
+ 2{\cal E}_s(\omega)p_{s}(2\omega)p_{s}(\omega) \right] 
+ S_3p_{s}(2\omega)p_{s}^2(\omega) 
\right).
\end{eqnarray}
Equation (15) is the main result of this section. 
The left hand side of Eq.(15) gives the effective SHG susceptibility of 
the composite 
{\em per unit volume} of the nonlinear coating material normalized 
by the SHG susceptibility of the coating material.  
Here $p_{s}(\omega)$ and $p_{s}(2\omega)$ 
are the dipole moments that would be induced 
in a {\em linear} particle when an electric field
is applied at frequencies $\omega$ or $2\omega$.  

The magnitude of the right hand side of Eq.(15) thus gives 
the possible enhancement in the effective SHG coefficient due 
to the geometry of the particles (coated particles) and the host 
medium through the factor $\Gamma_{s}$.  As a model system, we 
consider spherical particles in which the inner core is a 
Drude metal, which has a dielectric constant of the form 
\begin{equation}
\epsilon_\omega=1-\frac{\omega_p^2}{\omega^2+i\omega/\tau}
\end{equation}
with $\omega_p$ being the plasma frequency and $\tau$ 
the relaxation time.  We take 
$\omega_{p} = 2.28\times 10^{16} s^{-1}$ 
and $\tau=6.9\times 10^{-15}s$, numbers that 
correspond to bulk aluminum.  Note that 
$\omega_{p}\tau \sim 100 - 200$ are typical of metals.  
The nonlinear shell is assumed to have a frequency independent 
dielectric constant $\epsilon_{2} = 2.52$, while the host medium 
also has a frequency independent dielectric constant $\epsilon_{3}=1.76$. 
These numbers are typical of non-conducting materials. 
Figure 1 shows the real and imaginary parts of 
$d^{e}_{\omega,\omega}/(f d_{\omega,\omega})$ for this model system 
as a function of frequency $\omega/\omega_{p}$ and   
for different ratios of outer to inner radii.  We note that 
an enhancement factor of the order of $10^{3}$ can be achieved at 
suitable frequencies.  At certain resonance frequency of the coated 
particle, the local field in the shell region is sufficiently large to 
drive a much enhanced SHG susceptibility.  This frequency corresponds 
to the surface plasmon resonance of the coated particle, the value of 
which depends on the core material, the coating material, and the 
host medium.  The frequency can also be tuned by properly selecting 
the material parameters and/or by tuning the ratio of the outer to inner 
radii.  We see in Fig.1 that as the ratio increases, the resonance 
shifts to lower frequency. 
We also note that higher enhancement per unit volume of 
nonlinear coating material is achieved for thinner coating.  Therefore, 
one would expect a dilute suspension of spherical particles coated 
with a thin layer of nonlinear materials will give the largest 
enhancement per unit volume of nonlinear material.

\section{2D case: Coated cylinders}

A two dimensional random composites on the $x$-$y$ plane, for example, 
can be considered as a random 
dispersion of cylindrical particles in a host medium, with the axes of 
cylinders aligned in the same direction perpendicular to the plane.  A 
cross section of the system consists of circular 
discs embedded in the host. 
A calculation of the electric field ${\bf E}_{c}$ 
in the coated shell of a cylinder
in the presence of an applied field $E_{0}\hat{x}$
can be carried out in a way similar to the 3D case using standard 
electrostatics.  Instead of Eq.(10) in 3D, we have for the 2D case 
\begin{equation}
{\bf E}_{c} = {\cal E}_{c} \hat{x}+\frac{2(x^2\hat{x}+xy\hat{y})
-r^2\hat{x}}{r^4}p_c ,  
\end{equation}
where $r$ is the distance from the center of a disc (or cylinder) to 
the position under consideration inside the coating shell.  Here, 
\begin{eqnarray}
E_c & = & \Gamma_c E_0 ,  \\
p_c & = & \lambda_c r_{1}^2 \Gamma_{c} E_0 , \\
\Gamma_c & = & \frac{2\epsilon_3}{\epsilon_2+\epsilon_3+
(\epsilon_2-\epsilon_3)\lambda_c/\mu_c} , \\
\lambda_c & = & \frac{\epsilon_1-\epsilon_2}{\epsilon_1+\epsilon_2} , \\
\mu_c&=&(\frac{r_2}{r_1})^2, 
\end{eqnarray}
and the subscript ``$c$" is used to indicate that the case of coated cylinders
or discs are being considered. 
Comparing with results of a single dielectric circular disc in a host, a 
coated disc behaves as a disc with an effective 
dielectric constant $\tilde{\epsilon}_{c}$ of the form 
\begin{equation}
\tilde{\epsilon}_{c}=\frac{1+\lambda_{c}/\mu_{c}}
{1-\lambda_c/\mu_c}\epsilon_2 ,
\end{equation}
with $\lambda_{c}$ and $\mu_{c}$ given by Eqs.(21) and (22). 

To obtain an expression for 2D similar to Eq.(15), we make 
the same assumption for the SHG coefficient of the coating 
material.  This leads to an expression similar to Eq.(12) for the effective 
SHG suscepbility, except that we need to evaluate the average over 
the shell {\em area} of the quantity 
\begin{equation}
\left[{\cal E}_c(2\omega)+\frac{2x^2-r^2}{r^4}p_{c}(2\omega)\right]
\left[{\cal E}_c(\omega)+\frac{2x^2-r^2}{r^4}p_{c}(\omega)\right]^2, 
\end{equation} 
instead of the expression (13) in 3D. 
Defining $B_{c} = (2x^{2}-r^{2})/r^{4}$, 
the average is easily evaluated by noting that 
\begin{eqnarray}
\left\langle B_{c} \right\rangle_{shell} & = & 0 \nonumber \\
\left\langle B_{c}^{2} \right\rangle_{shell} & \equiv \tilde{S}_2 & = 
\frac{1}{a_{shell}}\left(\frac{1}{r_{1}^2}-\frac{1}{r_{2}^2}\right)
\frac{\pi}{2} \nonumber \\
\left\langle B_{c}^{3} \right\rangle_{shell} & = & 0. 
\end{eqnarray}
Here $a_{shell}$ is the area of the circular shell (ring) of coating. 

The effective SHG susceptibility is then found to be 
\begin{equation}
\frac{d^{e}_{\omega,\omega}}{f d_{\omega,\omega}}   
=\frac{1}{E_{0,2\omega}E_{0,\omega}^2}[{\cal E}_{c}(2\omega) 
{\cal E}_{c}^{2}(\omega) +\tilde{S}_2({\cal E}_c(2\omega) 
p_{c}^{2}(\omega) +
2{\cal E}_{c}(\omega) p_{c}(2\omega) p_{c}^{2}(\omega))]. 
\end{equation}
Equation (26) is the 
result in 2D, analogous to Eq.(15) in 3D.  It gives 
the effective SHG suscepbitility {\em per unit area} 
of nonlinear coating medium normalized to the SHG susceptibility 
of the coating medium in a 2D dilute 
suspension of coated discs in a linear host. 

Figure 2 shows $d^{e}_{\omega,\omega}/(f d_{\omega,\omega})$ 
as a function of frequency in the 2D case, for the same model system  
considered in the previous section.  
The features are similar to those in the 3D case.  
The enhancement factor tends to be larger in 
2D than in 3D for the same set of model parameters.  For discs 
coated with thin layer of nonlinear material, the SHG response 
per unit area of nonlinear material can be driven by the local 
field to achieve an enhancement of $10^{3}$, for modest choice of 
material parameters. 

\section{Discussion}
It is straightforward to generalize the formalism to third 
harmonic generation (THG)\cite{hui2}.  For coating nonlinear material in which 
the leading nonlinear response is THG with the THG coefficient 
denoted by $\chi_{\omega,\omega,\omega}$, we can again use the 
local electric field at the shell region to drive an enhanced 
THG response per unit volume of nonlinear material in a dilute suspension. 
The derivation is similar to that in Sec.II, although somewhat tedious. 
The result is 
\begin{eqnarray}
\frac{\chi^{e}_{\omega,\omega,\omega}}
{f \chi_{\omega,\omega,\omega}} & = &
\frac{1}{E_{0}(3\omega) E_{0}^{3}(\omega)} 
[{\cal E}_{s}(3\omega) {\cal E}_{s}^{3}(\omega) 
+ 3 S_{2} {\cal E}_{s}(\omega) p_{s}(\omega) 
({\cal E}_{s}(\omega) p_{s}(3\omega) + {\cal E}_{s}(3\omega) p_{s}(\omega)) + \nonumber \\
& & + S_3 p_{s}^{2}(\omega)(3 {\cal E}_{s}(\omega) p_{s}(3\omega) 
+ {\cal E}_{s}(3\omega) p_{s}(\omega)) 
+ S_4 p_{s}(3\omega) p_{s}^{3}(\omega))] , 
\end{eqnarray}
where 
\begin{equation} 
S_{4} = \left\langle B_{s}^{4} \right \rangle  
= \frac{1}{v_{shell}}(\frac{1}{r_1^9}-\frac{1}{r_2^9})\frac{64\pi}{105}
\end{equation}
for a dilute suspension of coated spherical particles.  

In the 2D case of coated discs of nonlinear THG coating material, 
the result is 
\begin{eqnarray}
\frac{\chi^{e}_{\omega,\omega,\omega}}{f \chi_{\omega,\omega,\omega}} &=&
\frac{1}{E_{0}(3\omega) E_{0}^{3}(\omega)}
[{\cal E}_{c}(3\omega) {\cal E}_{c}^{3}(\omega)
+ 3 \tilde{S}_2 {\cal E}_{c}(\omega) p_{c}(\omega) 
({\cal E}_{c}(\omega) p_{c}(3\omega) + {\cal E}_{c}(3\omega) p_{c}(\omega)) + \nonumber \\
&&+\tilde{S}_4 p_{c}(3\omega) p_{c}^{3}(\omega)] ,  
\end{eqnarray}
where 
\begin{equation}
\tilde{S}_{4} = \left\langle B_{c}^{4} \right\rangle = \frac{1}{a_{shell}}
(\frac{1}{r_{1}^{6}} - \frac{1}{r_{2}^{6}}) \frac{\pi}{8}. 
\end{equation}

To illustrate the enhancement in THG susceptibility, we use the same 
choice of material paramters as our model system and calculate the 
enhancement factor corresponding to Eq.(27) for spherical particles. 
Results are shown in Fig.3.  Since the THG is driven by higher power 
of the local field, the enhancement factor is found to be of the 
order of $10^{5}$ at resonance frequency, for our choice of parameters. 

Note that, when we calculate the effective SHG and THG susceptibilities
in the present work, we do so by treating the medium as effectively 
{\em homogeneous}, with effective properties.  Thus, our calculation
would lead to a collimated beam of harmonic emission when this
effective medium is used in an SHG experiment.  Besides this collimated 
emission, some diffuse scattering at the second-harmonic frequency is
also expected, because of the inhomogeneity of the medium.  Indeed,
such diffuse SHG scattering has been observed in a recent experiment,
carried out near the percolation threshold of a random metal-insulator
composite\cite{breit}, in addition to the collimated beam described here.  
This diffusive scattering should have an intensity 
varying as a power of the particle radius.  The present 
approach would need to be extended to include such diffuse scattering.  

Finally, we briefly comment on the validity of the
low-concentration approximation used to obtain eqs.\
(15), (16), (27) and (29).  These equations are obtained by calculating the 
fields and displacement vectors as if the coated particles feel 
a {\em uniform} applied field.  In actuality, the local field should 
include the dipole fields produced by the other particles.  
These fields may start to become significant when the dipole field due to one 
particle, calculated at the position of a neighboring particle, 
becomes comparable to the applied electric field.     
Although this condition may be satisfied
near resonance even for $f \ll 1$, the large enhancements we find may
still persist at higher $f$, because the interaction fields,
even though they can be substantial, may approximately cancel out if the 
environment of a particle can be approximated as isotropic.   The local field
factors at higher concentrations can be estimated using such methods as
the Maxwell-Garnett or effective-medium approximations.

Our results for SHG (Eqs.(15) and (26)) and THG (Eqs.(27) and (29)) 
are applicable to other possible structures of coated particles, as 
long as the composite is in the dilute concentration limit and
SHG originates from the coating material.
One could use a dielectric core and a metallic shell with a non-vanishing
SHG coefficient\cite{xu}. 
In general, careful choices can be made
so that the local electric field in the coating region is enhanced.  This
enhanced local field can then be used to drive the SHG response. 
In addition, it is possible to tune the local field by tuning 
the core size and the shell thickness.  
For composites of higher concentrations,
the Maxwell-Garnett approximation
and the effective-medium appoximation can be invoked in estimating
the local field factors.  In higher
concentrations, percolation effects can also lead to possible 
enhancement in SHG and other nonlinear response. 

\acknowledgments{This work was supported in part by a grant from the 
Research Grants Council of the Hong Kong SAR Government through 
the grant CUHK4129/98P.  One of us (DS) acknowledges the support
from the National Science Foundation through the grant DMR01-04987.}

\begin{figure}
\bigskip
\caption{The real and imaginary parts of the effective SHG
coefficient $d^{e}_{\omega,\omega}/(f d_{\omega,\omega})$
per unit volume of nonlinear
coating material as a function of frequency $\omega/\omega_{p}$
for a dilute suspension of particles with metallic core coated
with a nonlinear material.
Results for three different ratios of the outer to inner radii $\mu_{s}$
of the particles are shown.}
\bigskip
\bigskip
\caption{The real and imaginary parts of the effective SHG
coefficient $d^{e}_{\omega,\omega}/(f d_{\omega,\omega})$
per unit area of nonlinear
coating material as a function of frequency $\omega/\omega_{p}$
for a dilute suspension of discs with metallic core coated
with a nonlinear material in a 2D system. 
Results for three different ratios of the outer to inner radii $\mu_{c}$
of the discs are shown.}

\bigskip
\bigskip
\caption{The real and imaginary parts of the effective THG 
coefficient $\chi^{e}_{\omega,\omega,\omega}/(f \chi_{\omega,\omega,\omega})$
per unit volume of nonlinear
coating material as a function of frequency $\omega/\omega_{p}$
for a dilute suspension of particles with metallic core coated
with a nonlinear material.
Results for three different ratios of the outer to inner radii $\mu_{s}$
of the particles are shown.}

\end{figure}

\newpage



\newpage
\begin{thebibliography}{99}
\bibitem{bergman1} D. J. Bergman and D. Stroud,
Solid State Physics {\bf 46}, 147 (1992). 

\bibitem{shalaev1} V. M. Shalaev, Phys. Rep. {\bf 272}, 61 (1996).     

\bibitem{flytzanis} C. Flytzanis, F. Hache, M. C. Klein, and P. Roussignol, 
in {\it Progress in Optics} {\bf 29}, edited 
by E. Wolf (North-Holland, Amsterdam, 1991), p. 322.

\bibitem{liao} H. B. Liao, R. F. Xiao, J. S. Fu, H. Wang, K. S. Wong,
and G. K. L. Wong, Opt. Lett. {\bf 23}, 388 (1998).

\bibitem{klar} T. Klar, M. Perner, S. Grosse, G. von Plessen,
W. Spirkl, and J. Feldmann, Phys. Rev. Lett. {\bf 80}, 4249 (1998). 

\bibitem{jackson} J. B. Jackson, S. L. Westcott, L. R. Hirsch, J. L. West,
and N. J. Halas, Appl. Phys. Lett. {\bf 82}, 257 (2003).

\bibitem{averitt} R. D. Averitt, D. Sarkar, and N. J. Halas,
Phys. Rev. Lett. {\bf 78}, 4217 (1997).

\bibitem{zhou} H. S. Zhou, I. Honma, H. Komiyama,
and J.W. Haus, Phys. Rev. B {\bf 50}, 12052 (1994); M. Lomascolo,
A. Creta, G. Lao, L. Vasanelli, and L. Manna, Appl. Phys. Lett.
{\bf 82}, 418 (2003).

\bibitem{shalaev2} V. M. Shalaev, E. Y. Poliakov,
and V. A. Markel, Phys. Rev. B {\bf 53}, 2437 (1996); 
and references therein.

\bibitem{shalaev3} V. M. Shalaev, {\em Nonlinear Optics of Random Media: 
fractal composites and metal-dielectric films} (Springer, New York 2000).

\bibitem{brosseau} C. Brosseau and A. Beroual, Prog. Mat. Sci.
{\bf 48}, 373 (2003). 

\bibitem{shalaev4} V. M. Shalaev (ed.),
{\em Properties of nanostructured random media} (Springer, New York, 2002). 
\bibitem{stroud1} D. Stroud and P. M. Hui, Phys. Rev. B {\bf 37}, 8719 (1988).


\bibitem{stroud2} D. Stroud and Van E. Wood, J. Opt. Soc. Am. B {\bf 6}, 778 (1989).
\bibitem{neeves} A. E. Neeves and M. H. Birnboin,
J. Opt. Soc. Am. B {\bf 6}, 787 (1989);
Y. Q. Li, C. C. Sung, R. Inguva, and C. M. Bowden,
{\it ibid}. {\bf 6}, 814 (1989);
J. W. Haus, N. Kalyaniwalla, R. Inguva, M. Bloemer, and C. M. Bowden, {\it ibid}. {\bf 6}, 797 (1989).
\bibitem{zeng} X. C. Zeng, D. J. Bergman, P. M. Hui, and D. Stroud, Phys. Rev. B {\bf 38}, 10970 (1988).
\bibitem{haus} J. W. Haus, R. Inguva, and C. M. Bowden, Phys. Rev. A {\bf 40}, 5729 (1989).
\bibitem{bergman2} O. Levy,
D. J. Bergman, and D. G. Stroud, Phys. Rev. E {\bf 52}, 3184 (1995).
\bibitem{sipe} J. E. Sipe and R. W. Boyd, Phys. Rev. A {\bf 46},
1614 (1992); R. W. Boyd and
J. E. Sipe, J. Opt. Soc. Am. B {\bf 11}, 297 (1994).

\bibitem{hui1} P. M. Hui and D. Stroud, J. Appl. Phys. {\bf 82}, 4740 (1997).
\bibitem{hui2} P. M. Hui, P. Cheung, and D. Stroud, J. Appl. Phys. {\bf 84}, 3451 (1998).
\bibitem{butcher} P. N. Butcher and D. Cotter, {\em The Elements
of Nonlinear Optics} (Cambridge University Press, New York 1990). 
\bibitem{em} J. D. Jackson, {\em Classical Electrodynamics} 2nd edition,
(John Wiley \& Sons, New York 1975) Chapter 4.

\bibitem{breit} M. Breit, V. A. Podolskiy, S. Gr\'{e}sillon, G. von Plessen,
J. Feldmann, J. C. Rivoal, P. Gadenne, A. K. Sarychev, and V. M. Shalaev,
Phys. Rev. {\bf B64}, 125106 (2001).


\bibitem{xu} C. Xu and P. M. Hui, unpublished. 


\end{thebibliography}
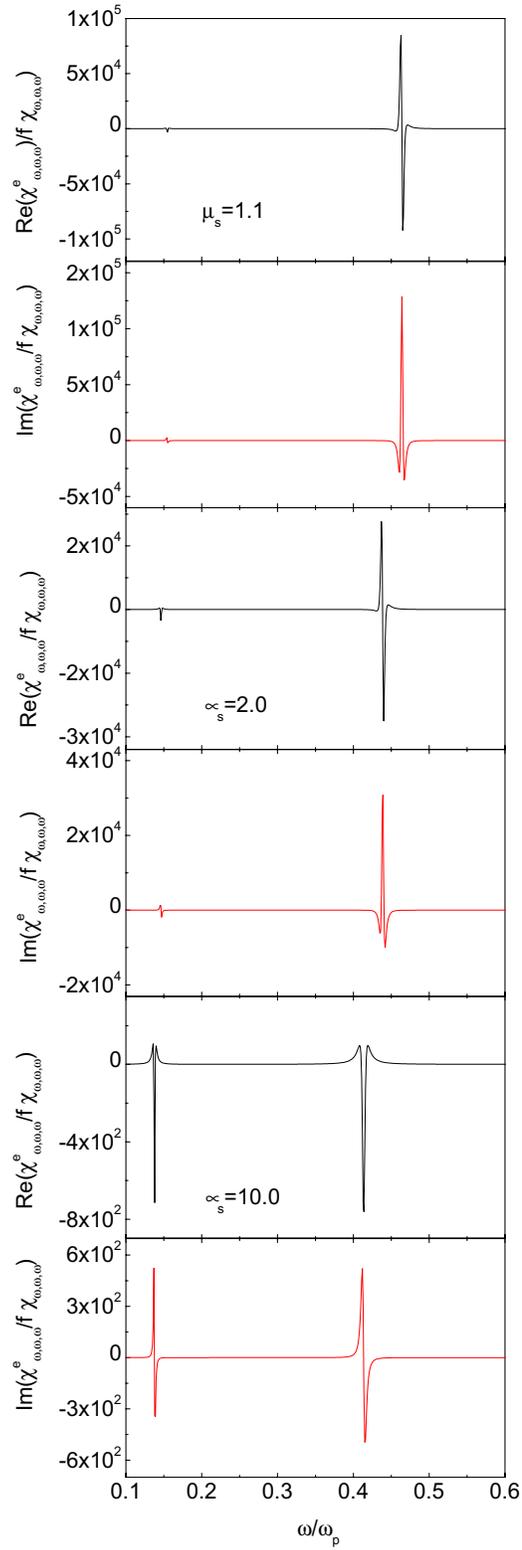
\end{document}